\pdfoutput=1

\documentclass[12pt,a4paper]{article}

\usepackage{ifthen} 
\newboolean{pdflatex}
\setboolean{pdflatex}{true} 
\usepackage{overpic}

\newboolean{articletitles}
\setboolean{articletitles}{true} 

\newboolean{uprightparticles}
\setboolean{uprightparticles}{false} 

\newboolean{inbibliography}
\setboolean{inbibliography}{false} 

\usepackage{lineno}  
\usepackage{xspace} 

\usepackage{graphicx}  
\usepackage{color}
\usepackage{colortbl}
\graphicspath{{./figs/}} 

\usepackage{amsmath} 
\usepackage{amssymb}
\usepackage{amsfonts}
\usepackage{upgreek} 

\newcommand*\patchAmsMathEnvironmentForLineno[1]{%
\expandafter\let\csname old#1\expandafter\endcsname\csname #1\endcsname
\expandafter\let\csname oldend#1\expandafter\endcsname\csname
end#1\endcsname
 \renewenvironment{#1}%
   {\linenomath\csname old#1\endcsname}%
   {\csname oldend#1\endcsname\endlinenomath}%
}
\newcommand*\patchBothAmsMathEnvironmentsForLineno[1]{%
  \patchAmsMathEnvironmentForLineno{#1}%
  \patchAmsMathEnvironmentForLineno{#1*}%
}
\AtBeginDocument{%
\patchBothAmsMathEnvironmentsForLineno{equation}%
\patchBothAmsMathEnvironmentsForLineno{align}%
\patchBothAmsMathEnvironmentsForLineno{flalign}%
\patchBothAmsMathEnvironmentsForLineno{alignat}%
\patchBothAmsMathEnvironmentsForLineno{gather}%
\patchBothAmsMathEnvironmentsForLineno{multline}%
}

\usepackage{hyperref}    
\usepackage[all]{hypcap} 

\def\lhcb {\mbox{LHCb}\xspace}

\ifthenelse{\boolean{uprightparticles}}%
{

 \def\Ppi         {\ensuremath{\uppi}\xspace}

 \def\Ppsi        {\ensuremath{\uppsi}\xspace}

 \def\PDelta      {\ensuremath{\Delta}\xspace}                 
 \def\PXi      {\ensuremath{\Xi}\xspace}                 
 \def\PLambda      {\ensuremath{\Lambda}\xspace}                 
 \def\PSigma      {\ensuremath{\Sigma}\xspace}                 
 \def\POmega      {\ensuremath{\Omega}\xspace}                 
 \def\PUpsilon      {\ensuremath{\Upsilon}\xspace}                 
 
 \def\PB      {\ensuremath{\mathrm{B}}\xspace}                 
                  
 \def\PD      {\ensuremath{\mathrm{D}}\xspace}

 \def\PJ      {\ensuremath{\mathrm{J}}\xspace}                 
 \def\PK      {\ensuremath{\mathrm{K}}\xspace}

 \def\Pb      {\ensuremath{\mathrm{b}}\xspace}                 
 \def\Pc      {\ensuremath{\mathrm{c}}\xspace}

 \def\Pi      {\ensuremath{\mathrm{i}}\xspace}

 \def\Ps      {\ensuremath{\mathrm{s}}\xspace}

}
{

 \def\Ppi         {\ensuremath{\pi}\xspace}

 \def\Ppsi        {\ensuremath{\psi}\xspace}                 
                  
 \mathchardef\PDelta="7101
 \mathchardef\PXi="7104
 \mathchardef\PLambda="7103
 \mathchardef\PSigma="7106
 \mathchardef\POmega="710A
 \mathchardef\PUpsilon="7107
                  
 \def\PB      {\ensuremath{B}\xspace}                 
                  
 \def\PD      {\ensuremath{D}\xspace}

 \def\PJ      {\ensuremath{J}\xspace}                 
 \def\PK      {\ensuremath{K}\xspace}

 \def\Pb      {\ensuremath{b}\xspace}                 
 \def\Pc      {\ensuremath{c}\xspace}

 \def\Pi      {\ensuremath{i}\xspace}

 \def\Ps      {\ensuremath{s}\xspace}

}

\def\squark    {\ensuremath{\Ps}\xspace}

\def\cquark    {\ensuremath{\Pc}\xspace}

\def\bquark    {\ensuremath{\Pb}\xspace}

\def\pion  {\ensuremath{\Ppi}\xspace}

\def\pim   {\ensuremath{\pion^-}\xspace}

\def\kaon  {\ensuremath{\PK}\xspace}
  \def\Kbar  {\kern 0.2em\overline{\kern -0.2em \PK}{}\xspace}

\def\Kp    {\ensuremath{\kaon^+}\xspace}
\def\Km    {\ensuremath{\kaon^-}\xspace}

  \def\Dbar    {\kern 0.2em\overline{\kern -0.2em \PD}{}\xspace}

\def\B       {\ensuremath{\PB}\xspace}
\def\Bbar    {\ensuremath{\kern 0.18em\overline{\kern -0.18em \PB}{}}\xspace}

\def\Bd      {\ensuremath{\B^0}\xspace}
\def\Bs      {\ensuremath{\B^0_\squark}\xspace}
\def\Bsb     {\ensuremath{\Bbar^0_\squark}\xspace}

\def\jpsi     {\ensuremath{{\PJ\mskip -3mu/\mskip -2mu\Ppsi\mskip 2mu}}\xspace}

  \def\Y#1S{\ensuremath{\PUpsilon{(#1S)}}\xspace}

\def\Lz {\ensuremath{\PLambda}\xspace}
\def\Lbar {\ensuremath{\kern 0.1em\overline{\kern -0.1em\PLambda}}\xspace}

\def\Lb      {\ensuremath{\Lz^0_\bquark}\xspace}

\def\to                 {\ensuremath{\rightarrow}\xspace}

\def\CP                {\ensuremath{C\!P}\xspace}

\newcommand{\dms}{\ensuremath{\Delta m_{\squark}}\xspace}

\newcommand{\DGs}{\ensuremath{\Delta\Gamma_{\squark}}\xspace}

\newcommand{\Gs}{\ensuremath{\Gamma_{\squark}}\xspace}

\newcommand{\phis}{\ensuremath{\phi_{\squark}}\xspace}
\newcommand{\betas}{\ensuremath{\beta_{\squark}}\xspace}

\def\AT#1     {\ensuremath{A_{\mathrm{T}}^{#1}}\xspace}           

\def\C#1      {\ensuremath{\mathcal{C}_{#1}}\xspace}                       
\def\Cp#1     {\ensuremath{\mathcal{C}_{#1}^{'}}\xspace}                    
\def\Ceff#1   {\ensuremath{\mathcal{C}_{#1}^{\mathrm{(eff)}}}\xspace}        
\def\Cpeff#1  {\ensuremath{\mathcal{C}_{#1}^{'\mathrm{(eff)}}}\xspace}       
\def\Ope#1    {\ensuremath{\mathcal{O}_{#1}}\xspace}                       
\def\Opep#1   {\ensuremath{\mathcal{O}_{#1}^{'}}\xspace}                    

\newcommand{\tev}{\ifthenelse{\boolean{inbibliography}}{\ensuremath{~T\kern -0.05em eV}\xspace}{\ensuremath{\mathrm{\,Te\kern -0.1em V}}\xspace}}
\newcommand{\gev}{\ensuremath{\mathrm{\,Ge\kern -0.1em V}}\xspace}
\newcommand{\mev}{\ensuremath{\mathrm{\,Me\kern -0.1em V}}\xspace}
\newcommand{\kev}{\ensuremath{\mathrm{\,ke\kern -0.1em V}}\xspace}
\newcommand{\ev}{\ensuremath{\mathrm{\,e\kern -0.1em V}}\xspace}
\newcommand{\gevc}{\ensuremath{{\mathrm{\,Ge\kern -0.1em V\!/}c}}\xspace}
\newcommand{\mevc}{\ensuremath{{\mathrm{\,Me\kern -0.1em V\!/}c}}\xspace}
\newcommand{\gevcc}{\ensuremath{{\mathrm{\,Ge\kern -0.1em V\!/}c^2}}\xspace}
\newcommand{\gevgevcccc}{\ensuremath{{\mathrm{\,Ge\kern -0.1em V^2\!/}c^4}}\xspace}
\newcommand{\mevcc}{\ensuremath{{\mathrm{\,Me\kern -0.1em V\!/}c^2}}\xspace}

\def\invfb   {\ensuremath{\mbox{\,fb}^{-1}}\xspace}

\def\fs   {\ensuremath{\rm \,fs}\xspace}

\def\invps{\ensuremath{{\rm \,ps^{-1}}}\xspace}

\def\gsim{{~\raise.15em\hbox{$>$}\kern-.85em
          \lower.35em\hbox{$\sim$}~}\xspace}
\def\lsim{{~\raise.15em\hbox{$<$}\kern-.85em
          \lower.35em\hbox{$\sim$}~}\xspace}

\def\rad{\ensuremath{\rm \,rad}\xspace}

\def\tell1  {TELL1\xspace}
\def\ukl1   {UKL1\xspace}

\usepackage{cite} 
\usepackage{mciteplus}

\def\BtoJpsiKK{B_s^0 \to J/\psi K^+K^-}
\def\BtoJpsipipi{B_s^0 \to J/\psi \pi^+\pi^-}

\def\delpar{\delta_\parallel}
\def\delperp{\delta_\perp}
\def\delzero{\delta_0}

\def\aparsq{|A_{\|}|^2}
\def\aperpsq{|A_{\perp}|^2}

\def\azerosq{|A_{0}|^2}

\def\maglambda{|\lambda|}

\def\phiszero{\phis^0}
\def\phispara{\phis^\parallel}
\def\phisperp{\phis^\perp}

\begin{document}

\renewcommand{\thefootnote}{\fnsymbol{footnote}}
\setcounter{footnote}{1}


\begin{titlepage}
\pagenumbering{roman}

\vspace*{-1.5cm}
\centerline{\large EUROPEAN ORGANIZATION FOR NUCLEAR RESEARCH (CERN)}
\vspace*{1.5cm}
\hspace*{-0.5cm}
\begin{tabular*}{\linewidth}{lc@{\extracolsep{\fill}}r}
\ifthenelse{\boolean{pdflatex}}
{\vspace*{-2.7cm}\mbox{\!\!\!\includegraphics[width=.14\textwidth]{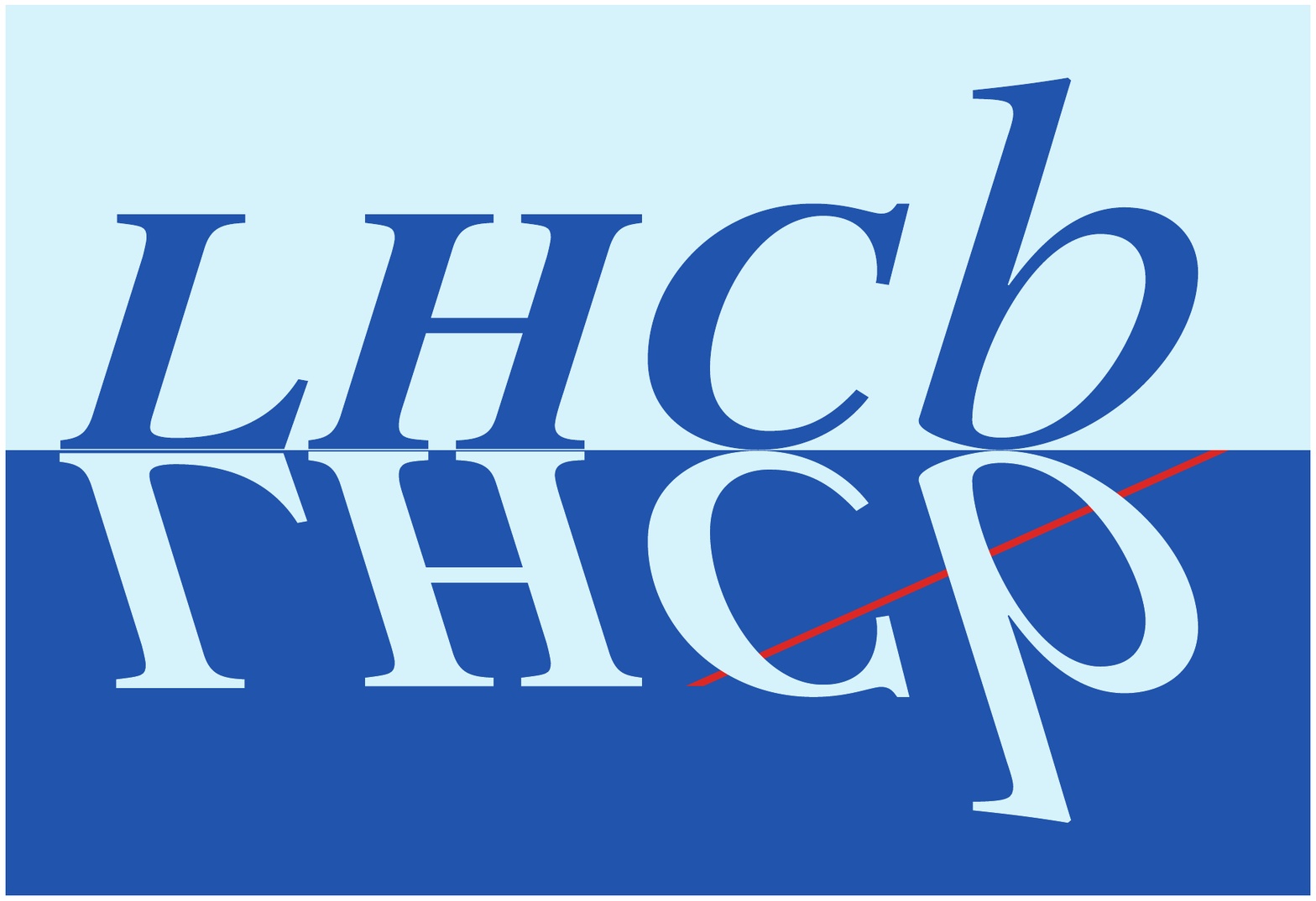}} & &}%
{\vspace*{-1.2cm}\mbox{\!\!\!\includegraphics[width=.12\textwidth]{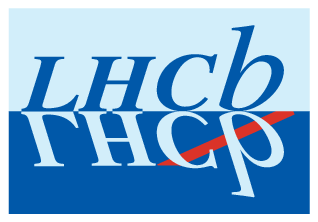}} & &}%
\\
 & & CERN-PH-EP-2014-271 \\  
 & & LHCb-PAPER-2014-059 \\  
 & & November 11, 2014\\ 
 & & \\
\end{tabular*}


{\bf\boldmath\huge
\begin{center}
Precision measurement of \CP violation in $\Bs \to \jpsi \Kp\Km$ decays
\end{center}
}

\vspace*{.0cm}

\begin{center}
The LHCb collaboration\footnote{Authors are listed at the end of this Letter.}
\end{center}


\begin{abstract}
  \noindent
  The time-dependent \CP\ asymmetry in  $\Bs\to J/\psi K^{+}K^{-}$ decays is measured
  using $pp$ collision data, corresponding to an integrated
  luminosity of $3.0$\invfb, collected with the LHCb detector at centre-of-mass
  energies of $7$ and $8$\,\tev. In a sample of 96 000
  $\Bs\to J/\psi K^{+}K^{-}$ decays, the \CP-violating phase \phis is measured, as well as the decay widths
  $\Gamma_{\mathrm{L}}$ and $\Gamma_{\mathrm{H}}$ of the light and
  heavy mass eigenstates of the \Bs--\Bsb system. The values obtained are
  $\phi_s = -0.058 \pm 0.049 \pm 0.006$\rad, 
  $\Gamma_s \equiv (\Gamma_{\mathrm{L}}+\Gamma_{\mathrm{H}})/2 =  0.6603  \pm  0.0027  \pm  0.0015$\invps, 
  and
  $\DGs   \equiv \Gamma_{\mathrm{L}} -
  \Gamma_{\mathrm{H}}  = 0.0805  \pm 0.0091 \pm  0.0032$ \invps,
where the first uncertainty is statistical and the second systematic. These are the  most precise single measurements of those quantities to date.
A combined analysis with $B_s^{0} \rightarrow J/\psi \pi^{+}\pi^{-}$ decays gives
\mbox{$\phi_s = -0.010  \pm  0.039 $\rad}.
All measurements are in agreement with the Standard Model predictions.
For the first time the phase $\phis$ is measured independently for each
polarisation state of the $K^{+}K^{-} $ system and shows no evidence
for polarisation dependence.
\end{abstract}

\vspace*{.0cm}

\begin{center}
  Submitted to Phys. Rev. Lett.
\end{center}

\vspace{\fill}

{\footnotesize 
    \centerline{\copyright~CERN on behalf of the \lhcb collaboration, license \href{http://creativecommons.org/licenses/by/4.0/}{CC-BY-4.0}.}}
\vspace*{2mm}

\end{titlepage}


\newpage
\setcounter{page}{2}
\mbox{~}

\cleardoublepage


\renewcommand{\thefootnote}{\arabic{footnote}}
\setcounter{footnote}{0}



\pagestyle{plain} 
\setcounter{page}{1}
\pagenumbering{arabic}


%


The \CP-violating phase \phis arises in the interference between the amplitudes of
\Bs mesons decaying via $b\to c\overline{c}s$ transitions to \CP eigenstates directly and those decaying after oscillation. 
In the Standard Model (SM), ignoring sub-leading contributions, this phase is predicted to be $-2\betas$, where $\betas=\arg\left[                                 
 - (V_{ts} V_{tb}^*) / (V_{cs} V_{cb}^*)\right]$ and $V_{ij}$ are elements of the quark-mixing
 matrix~\cite{Kobayashi:1973fv,*Cabibbo:1963yz}. 
Global fits to experimental data give
 \mbox{$-2\betas=-0.0363\pm0.0013\rad$}~\cite{CKMfitter}.
This phase could be modified if non-SM particles were to contribute to the \Bs--\Bsb\ oscillations~\cite{Buras:2009if,Chiang:2009ev} 
and a measurement of $\phis$ significantly different from the SM prediction would provide unambiguous evidence for
processes beyond the SM. 

The LHCb collaboration has previously reported measurements of \phis 
using \mbox{$\BtoJpsiKK$} and $\BtoJpsipipi$ decays~\cite{cLHCb-PAPER-2011-021,LHCb-PAPER-2013-002}.
These measurements were based upon data, corresponding to an integrated luminosity of up to $1.0\invfb$,
collected in {\it pp}
collisions at a centre-of-mass energy of  $7\tev$ in 2011 at the LHC.
The ATLAS, CDF and D0 collaborations have also measured \phis in \mbox{$\BtoJpsiKK$} decays~\cite{Aad:2014cqa,Aaltonen:2012ie,Abazov:2011ry}.
This Letter extends the LHCb measurements in the \mbox{$\BtoJpsiKK$} channel by
adding data corresponding to $2.0\invfb$ of integrated luminosity collected at $8\tev$ in 2012,  
and presents the combined results for \phis  including the analysis of $\BtoJpsipipi$ decays
from Ref.~\cite{LHCb-PAPER-2014-019}.
For the first time, the \CP-violating phases are measured separately for
each polarisation state of the $\Kp\Km$ system. Knowledge of these parameters
is an important step towards the control of loop-induced effects to the decay amplitude,
which could potentially be confused with non-SM contributions to \Bs--\Bsb mixing~\cite{Faller:2008gt}.
The analysis of the  \mbox{$\BtoJpsiKK$} channel reported here is as
described in Ref.~\cite{LHCb-PAPER-2013-002},
to which the reader is referred for details,
except for the changes described below.

The \lhcb detector is a single-arm forward
spectrometer covering the \mbox{pseudorapidity} range $2<\eta <5$,
designed for the study of particles containing \bquark or \cquark
quarks, and is described in Ref.~\cite{Alves:2008zz}.
The trigger~\cite{LHCb-DP-2012-004} consists of a hardware stage, based on information from the 
calorimeter and muon systems, followed by a software stage, in which all charged particles
with transverse momentum greater than 500\,(300)\mevc are reconstructed for 2011 (2012) data.
Further selection requirements are applied offline, as described in Ref.~\cite{LHCb-PAPER-2013-002},
in order to increase the signal purity.

The $\Bs\to \jpsi\Kp\Km$ decay proceeds
predominantly via $\Bs\to \jpsi\phi$, in which the $\Kp\Km$ pair
from the $\phi$ is in a P-wave configuration.
The final state is a superposition of \CP-even and \CP-odd
states depending upon the relative orbital angular momentum of the $\jpsi$ and $\phi$ mesons.
The $\jpsi\Kp\Km$ final state can also be produced with $\Kp\Km$ pairs in a \CP-odd
S-wave configuration~\cite{Stone:2008ak}. The measurement of 
$\phis$ requires the \CP-even and \CP-odd components to
be disentangled by analysing the distribution of the reconstructed decay angles
of the final-state particles.
In this analysis  the decay angles are defined in the helicity basis, 
$\cos\theta_K$, $\cos\theta_\mu$, and $\varphi_h$, as described in Ref.~\cite{LHCb-PAPER-2013-002}.

\begin{figure}[t]
\begin{center}
\begin{tabular}{cc}
  \begin{overpic}[width=0.49\textwidth]{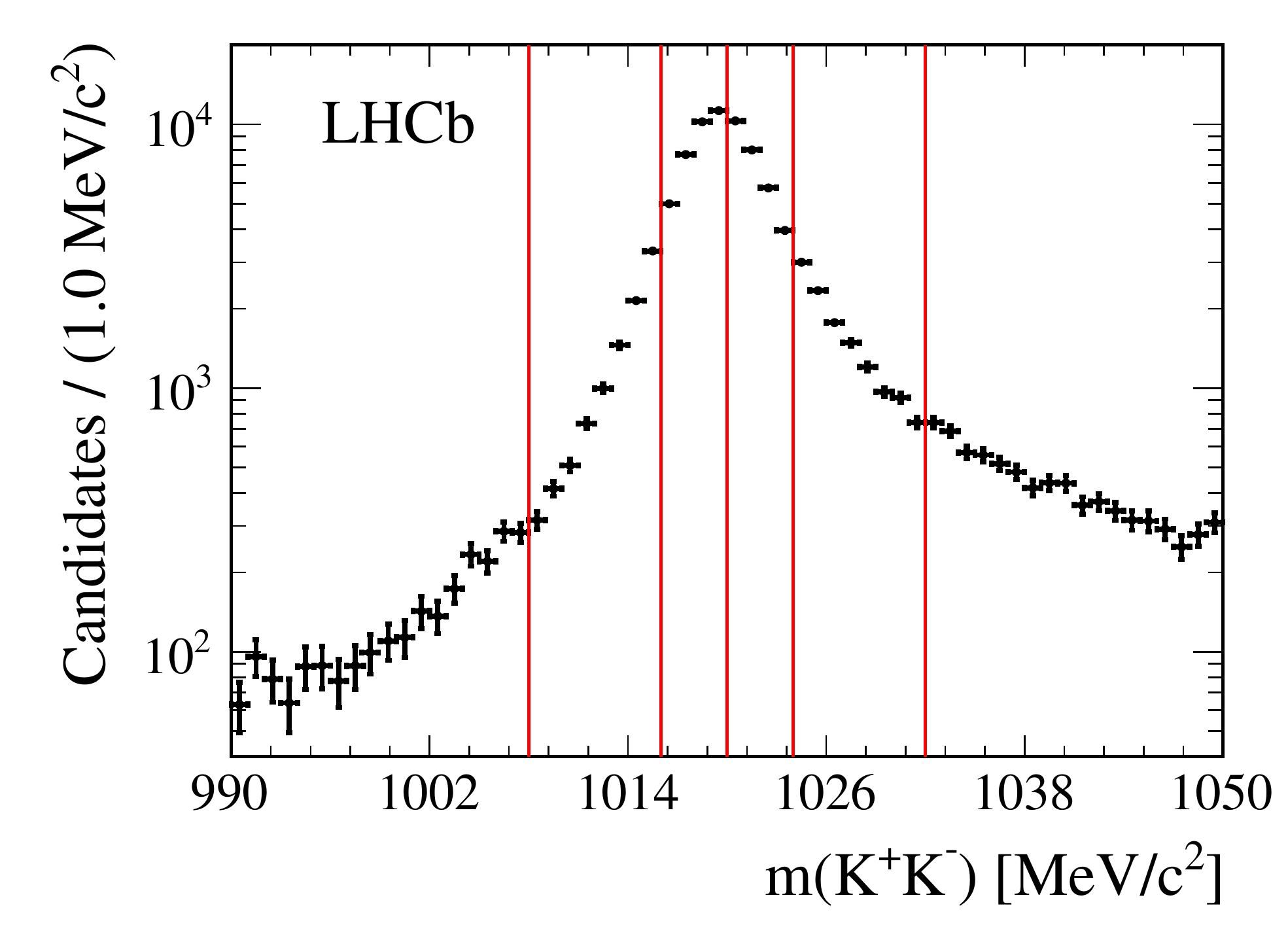}
    \put(26,46){(a)}
  \end{overpic}
  &
  \begin{overpic}[width=0.49\textwidth]{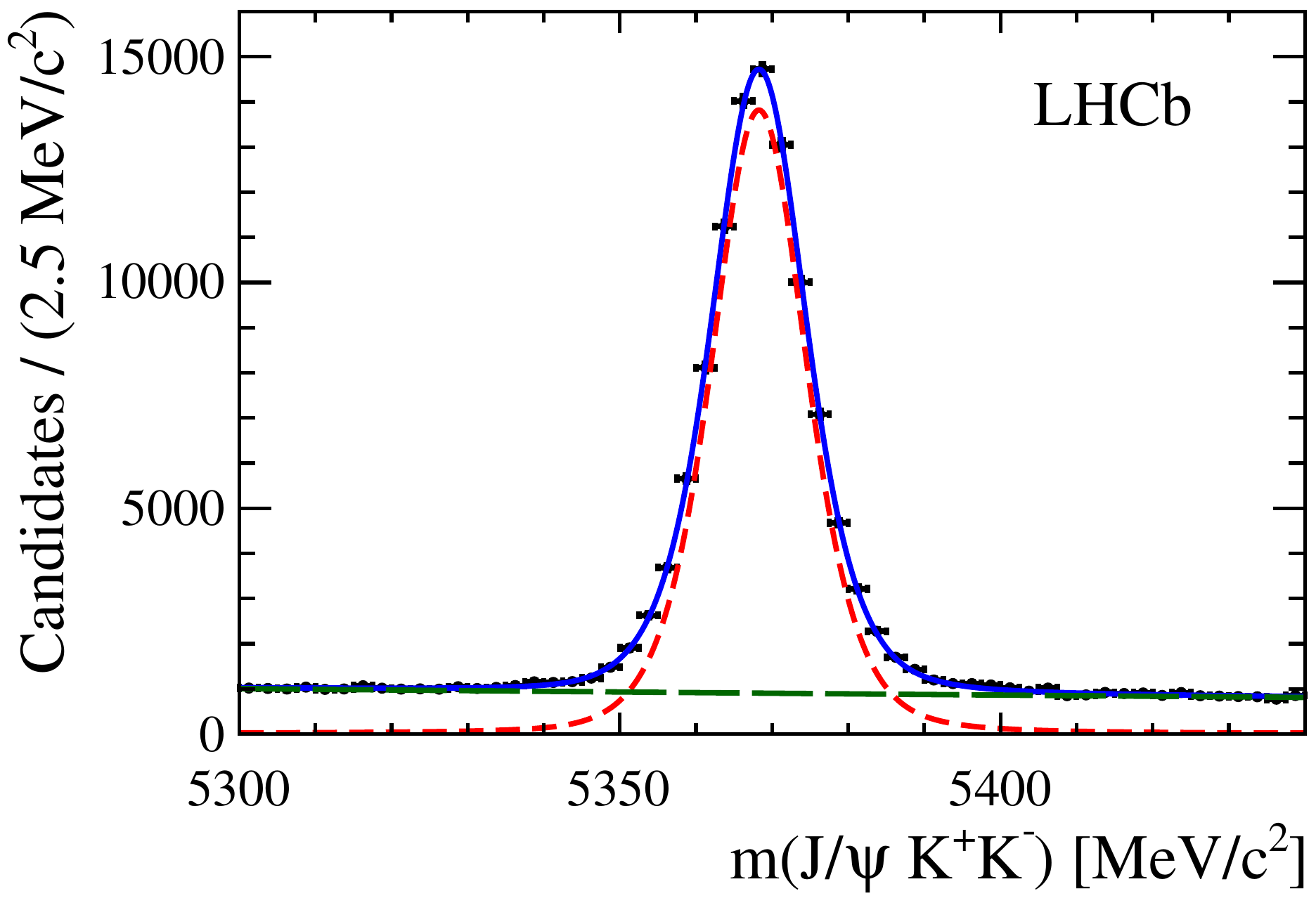}
    \put(26,46){(b)}
  \end{overpic}
  \end{tabular}
  \caption{\small (a) Background-subtracted invariant mass distributions of the $K^{+}K^{-}$ system in the
	   selected $\Bs\to\jpsi K^{+}K^{-}$ candidates (black points).
	   The vertical red lines denote the boundaries of the six bins used in the maximum likelihood fit.
	   (b) Distribution of $m(\jpsi\Kp\Km)$ for the data sample (black points) and projection of the maximum likelihood fit
	   (blue line).
The \Bs\ signal component is shown by the red dashed line and the combinatorial background by the green long-dashed line. 
Background from misidentified \Bd\ and \Lb\ decays is subtracted, as described in the text.}
\label{fig:mass}
\end{center}
\end{figure}

The invariant mass distributions for $\Kp\Km$ and $\jpsi(\to\mu^+\mu^-) \Kp \Km$ candidates are shown 
in Figs.~\ref{fig:mass}(a) and~\ref{fig:mass}(b), respectively.
The combinatorial background is modelled with an exponential function 
and the \Bs\ signal shape is parameterised by a double-sided
Hypatia function~\cite{Santos:2013gra}. 
The fitted signal yield is $95\,690\pm350$.
In addition to the combinatorial background, studies of the
data in sidebands of the $m(\jpsi\Kp\Km)$ spectrum show 
contributions from approximately 1700 $B^0\to\jpsi\Kp\pim$ (4800 $\Lb\to\jpsi p\Km$)
decays where the pion (proton) is misidentified as a kaon.
These background events have complicated correlations between the angular variables  and  $m(\jpsi\Kp\Km)$.
In order to avoid the need to describe explicitly such correlations in the analysis, 
the contributions from these backgrounds are statistically subtracted by adding to the data
simulated events of these decays with negative weight.
Prior to injection, the simulated  events are weighted such that 
the distributions of the relevant variables used in the fit, and their correlations, match those of data.

The principal physics parameters of interest are  
$\Gs$,  $\DGs$, $\phis$, $\maglambda$, the \Bs\ mass difference, \dms,
and the polarisation amplitudes $A_k = |A_k|e^{-i\delta_k}$, 
where the indices $k\in\{0, \parallel, \perp, {\rm S}\}$ refer to the different polarisation states of the
$\Kp\Km$ system.
The sum $\aparsq+\azerosq+\aperpsq$ equals unity and by convention $\delzero$ is zero. 
The parameter $\lambda$ describes \CP{} violation in the interference between
mixing and decay and is defined by $\eta_k (q/p) (\bar{A}_k/A_k)$, where it is 
assumed to be the same for all polarisation states.
The complex parameters $p = \langle\Bs | B_{\rm L}\rangle$ and $q = \langle\Bsb | B_{\rm L}\rangle$
describe the relation between mass and flavour eigenstates and  $\eta_k$ is the \CP{} eigenvalue
of the polarisation state $k$. 
The \CP-violating phase is defined by $\phis \equiv - \arg{\lambda}$.
In the absence of \CP{} violation in decay, $\maglambda=1$. 
\CP violation in \Bs-meson mixing is negligible, following measurements in Ref.~\cite{LHCb-PAPER-2013-033}.
Measurements of the above parameters are obtained from a weighted maximum likelihood fit~\cite{sFit}
to the decay-time and helicity angle distributions of the data as described in Ref.~\cite{LHCb-PAPER-2013-002}.


The  \Bs decay-time distribution is distorted by the trigger selection requirements
and by the track reconstruction algorithms. Corrections are determined
from data using the methods described in Ref.~\cite{LHCb-PAPER-2013-065}
and are incorporated in the maximum likelihood
fit by a parameterised function, in the case of the trigger, and by per-candidate weights,
in the case of the track reconstruction. Both corrections are
validated using a sample of $10^6$ simulated $\Bs\to\jpsi\phi$ events.


To account for the experimental decay-time resolution, the signal 
probability density function (PDF) is defined per candidate and
is convolved with the sum of two Gaussian functions with
a common mean, $\mu$, and independent widths $w_i$, $i\in\{1,2\}$. The widths are given by
\mbox{$w_i \equiv r_i\sigma + s_i\sigma^2$}, where $r_i$ and $s_i$ are 
scale factors for each Gaussian function and $\sigma$ is the
per candidate decay-time uncertainty, estimated by the kinematic fit used to calculate
the decay time.
The scale factors are determined from the decay-time distribution of a control sample of
$\jpsi\Kp\Km$ candidates that are selected as for signal except for
decay-time requirements.
The average value of the $\sigma$ distribution in the sample of prompt candidates
is approximately $35$\fs and the effective average resolution is $46$\fs.


The flavour of the $\Bs$ candidate at production is inferred using two independent
classes of flavour tagging algorithms, the opposite-side (OS) tagger and the
same-side kaon (SSK) tagger, which exploit specific features of the production of $b\bar{b}$ quark pairs in $pp$ collisions. 
The OS tagger algorithm is described in Ref.~\cite{LHCb-PAPER-2013-002} but
is re-calibrated using data sets of flavour-specific decays,
yielding a tagging power of $(1.19\pm 0.06)\%$ for events with only an OS-tag.
The SSK algorithm deduces the signal production flavour by exploiting
charge-flavour correlations of the kaons produced during the
hadronisation process of the $\overline{b}$ quark forming the signal \Bs meson.
The tagging kaon is identified using a selection based on a neural
network that gives an effective tagging power of $(0.84\pm0.11)\%$, corresponding
approximately to a $40\%$ improvement in tagging power with respect to that
in Refs.~\cite{LHCb-PAPER-2013-002,LHCb-CONF-2012-033}. 
The SSK algorithm is calibrated using a sample of $\Bs\to D_s^- \pi^+$ decays.
For events that have both 
OS and SSK tagging decisions, corresponding to $26\%$ of the tagged sample,
the effective tagging power is $(1.70\pm0.08)\%$. The combined tagging power of
the three independent tagging categories defined above is $(3.73\pm0.15)\%$.

Due to different $m(\Kp\Km)$ line shapes of the S- and P-wave contributions, their interferences
are suppressed by an effective coupling factor after integrating over a finite 
$m(\Kp\Km)$ region.
The fit is carried out in six bins of $m(\Kp\Km)$,
as shown in Fig.~\ref{fig:mass}(a),
to allow measurement of the small S-wave amplitude in each bin and to
minimise correction factors in the interference terms of the PDF. 

\begin{figure}

\centering{
  \includegraphics[width=0.49\textwidth]{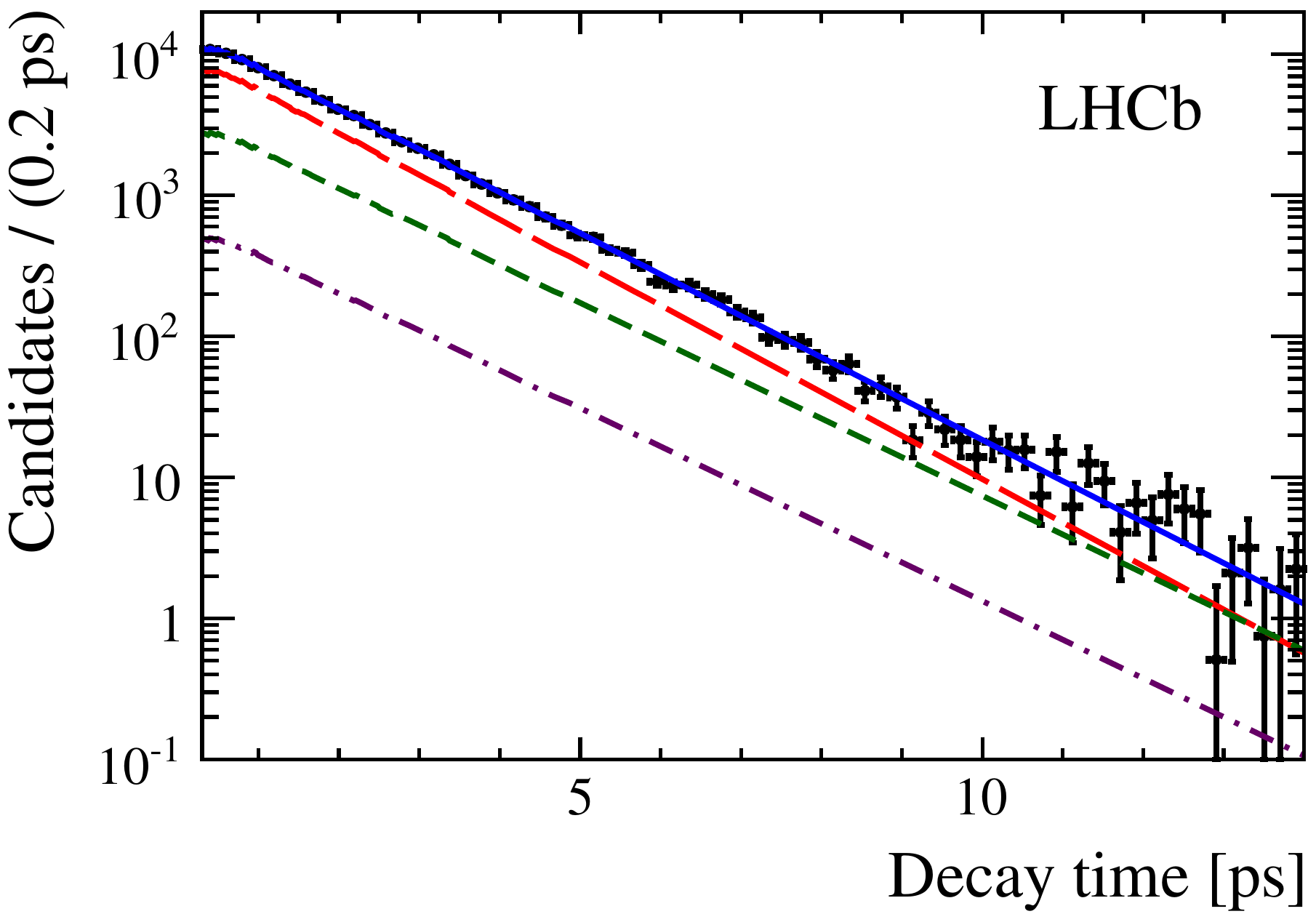}
  \includegraphics[width=0.49\textwidth]{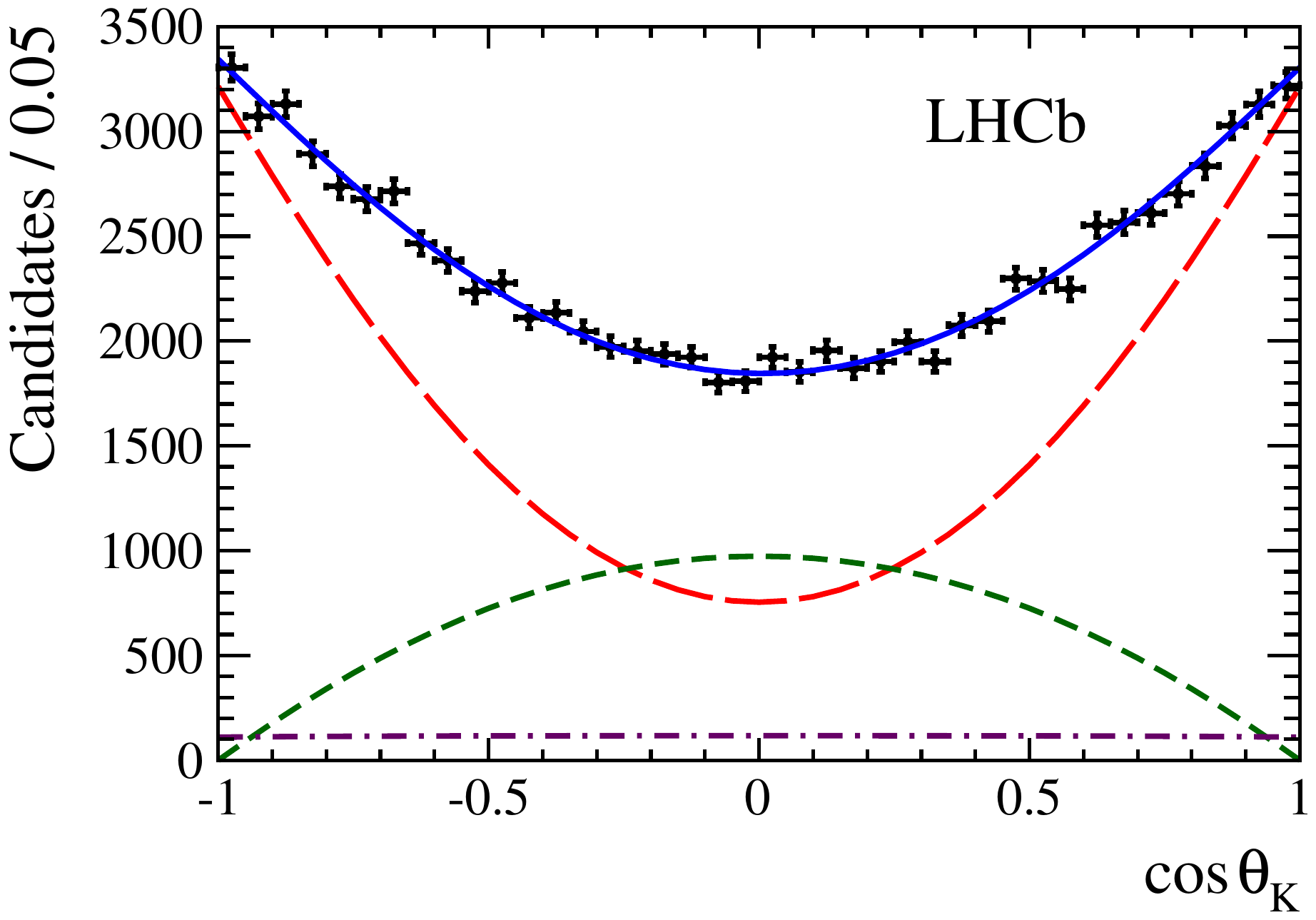}\\
  \includegraphics[width=0.49\textwidth]{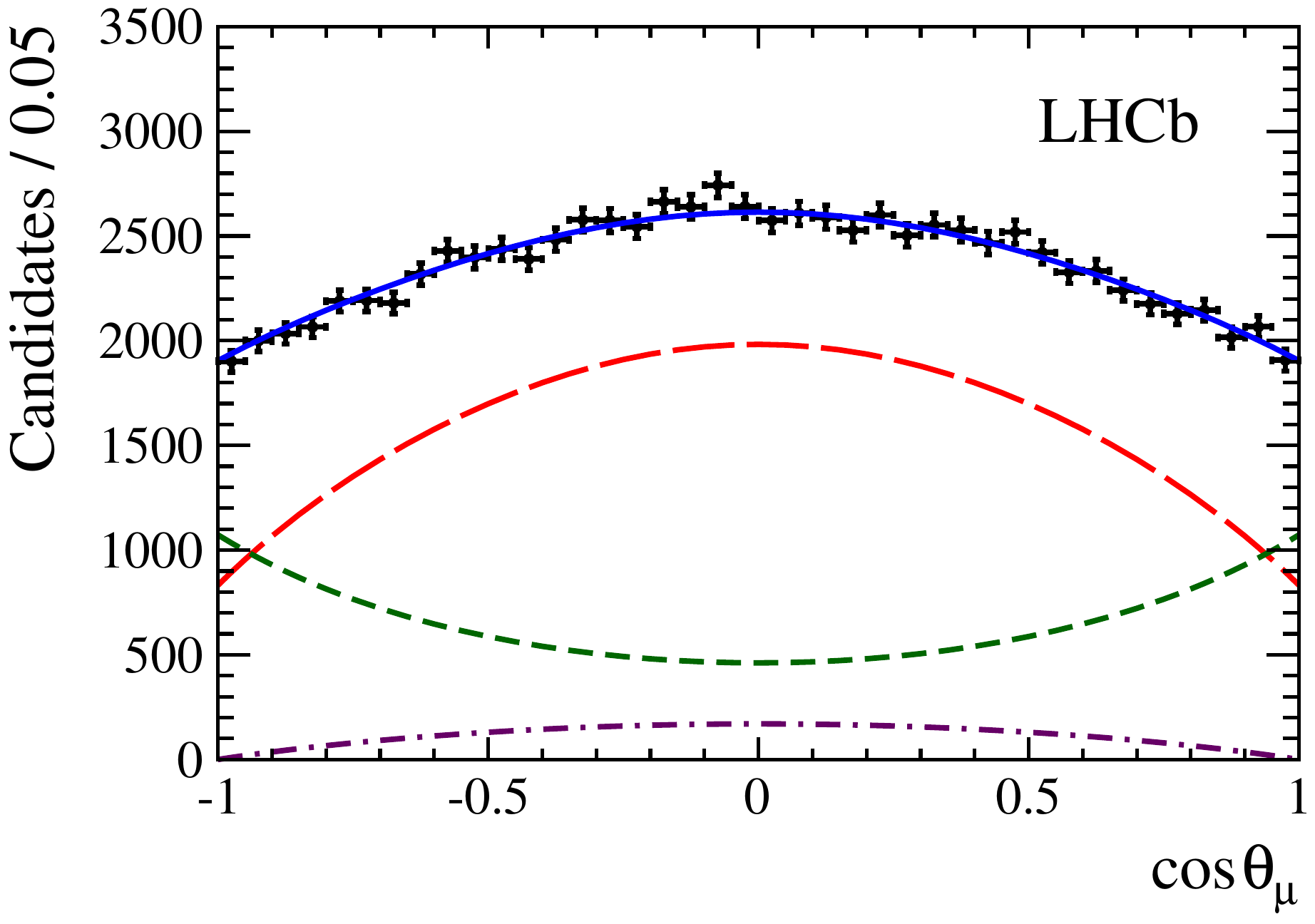}
  \includegraphics[width=0.49\textwidth]{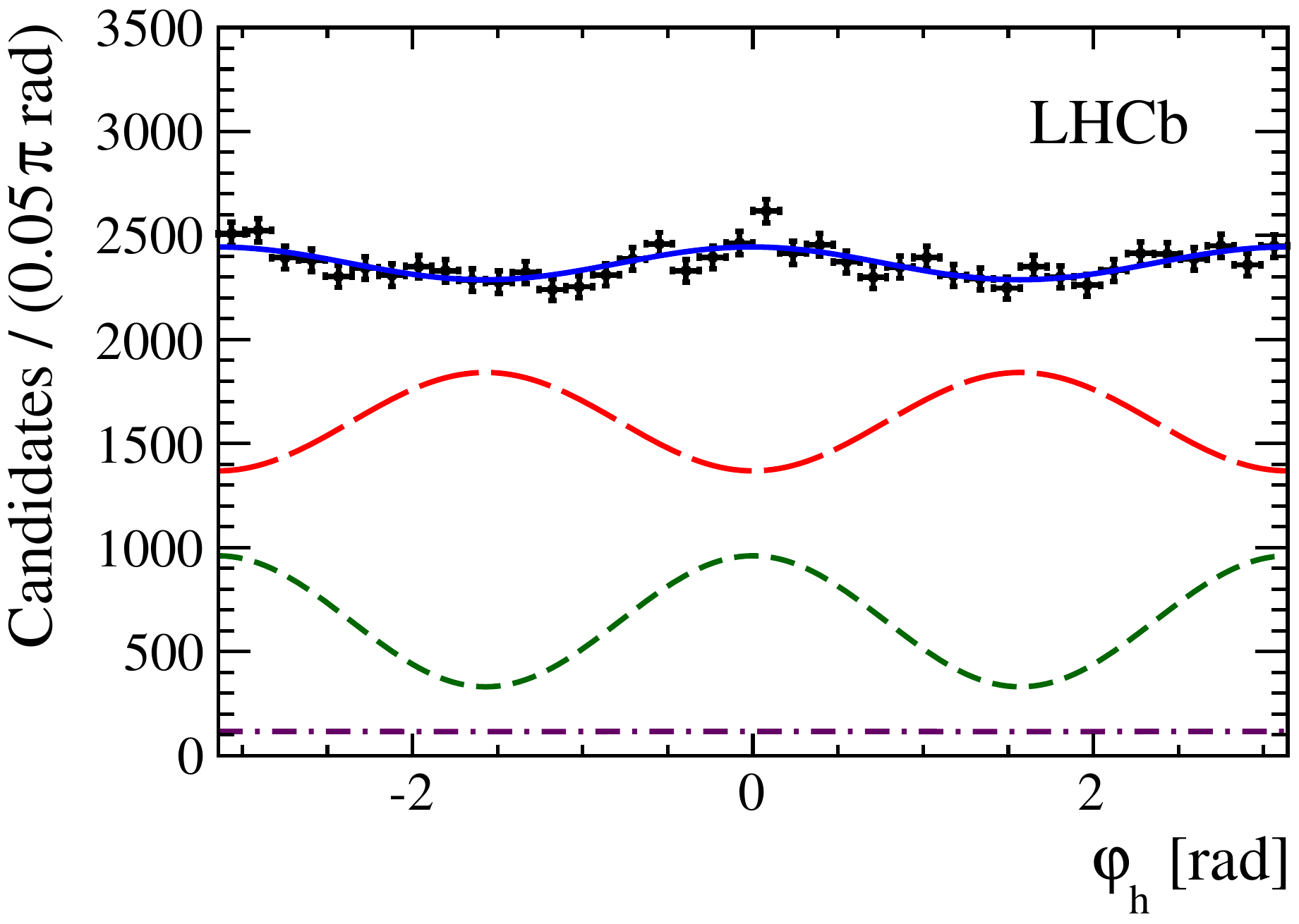}
}
\caption{\small\label{fig:results-projections} 
Decay-time and helicity-angle distributions for $\Bs\to\jpsi\Kp\Km$ decays (data points) with the one-dimensional fit projections overlaid. 
The solid blue line shows the total signal contribution, which is composed of 
\CP-even (long-dashed red), \CP-odd (short-dashed green) and S-wave (dotted-dashed purple) contributions.}
\end{figure}

The results of the fit are consistent with the measurements reported in
Ref.~\cite{LHCb-PAPER-2013-002} and are reported in Table~\ref{tab:results}
where the first uncertainty is statistical and the second systematic.
The correlation matrix is given in Ref.~\cite{supplementary}. 
In contrast to Ref.~\cite{LHCb-PAPER-2013-002} the value of $\dms$ is unconstrained in this
fit, thereby providing an independent measurement of this quantity,
which is consistent with the results of Ref.~\cite{LHCb-PAPER-2013-006}. 
The projections of the decay time and angular distributions are shown in Fig.~\ref{fig:results-projections}.

The results reported in Table~\ref{tab:results} are obtained with the assumption that $\phis$ and $\maglambda$
are independent of the final-state polarisation. This condition can be relaxed to allow the measurement of
$\phi_s^{k}$ and $|\lambda^k|$ separately for each polarisation,
following the formalism in Ref.~\cite{Liu:2013nea}.
The results of this fit are shown in Table~\ref{tab:results-phipol} and the statistical
correlation matrix is given in Ref.~\cite{supplementary}. 
There is no evidence for a polarisation-dependent \CP violation arising in $\Bs\to\jpsi\Kp\Km$ decays.
 
\begin{table}[t]
\caption{\small Values of the principal physics parameters determined from the polarisation-independent fit.}
\centerline{
    \begin{tabular}{l c}
                                Parameter    &     Value\\
      \hline
            \rule{0mm}{4.5mm}                          $\Gs$ $[\rm ps^{-1}]$    		&   $0.6603 \pm 0.0027        \pm  0.0015$  \\ 
      \rule{0mm}{4.5mm}                                $\DGs$ $[\rm ps^{-1}]$   	&   $0.0805  \pm 0.0091         \pm  0.0032$   \\
            \rule{0mm}{4.5mm}                          $\aperpsq$       		&   $0.2504 \pm 0.0049         \pm  0.0036$  \\
            \rule{0mm}{4.5mm}                          $\azerosq$        		&   $0.5241 \pm 0.0034          \pm  0.0067$  \\
             \rule{0mm}{4.5mm}                         $\delpar$~[rad]   	&   $3.26	\   ^{+0.10\ +0.06}_{-0.17\ -0.07}\ \ \, $    \\
              \rule{0mm}{4.5mm}                        $\delperp$~[rad]  	&   $3.08  \   ^{+0.14}_{-0.15}   \pm 0.06$    \\
                   \rule{0mm}{4.5mm}                   $\phis$~[rad]      	&   $-0.058   \pm 0.049              \pm  0.006$   \\
             \rule{0mm}{4.5mm}                         $\maglambda$    	&   $\phantom{+}0.964    \pm  0.019           \pm  0.007$   \\ 
            \rule{0mm}{4.5mm}                          $\dms$   $[\rm ps^{-1}]$  	&   $17.711\  ^{+0.055}_{-0.057} \pm0.011$   \\ 
            \hline
    \end{tabular}
     }
\label{tab:results}
\end{table}

\begin{table}[t]
\caption{\small Values of the polarisation-dependent parameters $\phi_s^k$ and $|\lambda^k|$ determined from the polarisation-dependent fit.}
\centerline{
    \begin{tabular}{l c}
                                Parameter    &     Value\\
      \hline
                                $|\lambda^{0}|$    	&   $\phantom{+}1.012      \pm  0.058                \pm  0.013$   \\ 
                                $|\lambda^{\parallel}/\lambda^{0}|$    	&   $\phantom{+}1.02\phantom{0} \pm 0.12\phantom{0} \pm 0.05\phantom{0}$   \\ 
                                $|\lambda^{\perp}/\lambda^{0}|$    	&   $\phantom{+}0.97\phantom{0} \pm 0.16\phantom{0} \pm 0.01\phantom{0}$   \\ 
			       $|\lambda^{\rm S}/\lambda^{0}|$        & $\phantom{+}0.86\phantom{0} \pm  0.12\phantom{0}  \pm 0.04\phantom{0}$ \\
                                $\phiszero$~[rad]      	&   $-0.045     \pm  0.053               \pm  0.007$   \\
                                $\phispara-\phiszero$~[rad]      	&   $-0.018     \pm  0.043               \pm  0.009$   \\
                                $\phisperp-\phiszero$~[rad]      	&   $-0.014     \pm  0.035               \pm  0.006$   \\
			       $\phi_s^{\rm S}-\phi_s^{0}$~[rad]  	& $\phantom{+}0.015     \pm  0.061 \pm 0.021 $ \\ 
                                \hline
    \end{tabular}
     }
\label{tab:results-phipol}
\end{table}

{
\begin{table}[h]
  \caption{\small Statistical and systematic uncertainties for the polarisation-independent result.}

  \begin{center}\footnotesize
    \begin{tabular}{lccccccccc}
      Source            			& $\Gs$ 	& $\DGs$	& $\aperpsq$ 	& $\azerosq$ 	& $\delpar$ 	&$\delperp$	& $\phi_s$  	& $\maglambda$  & $\dms$ 	        \\
       						& [ps$^{-1}$]	&[ps$^{-1}$]	&  		& 		&  [rad] 	&  [rad] 	& [rad] 	&	          &[ps$^{-1}$]      	\\[0.5ex] \hline
      \rule[3.5mm]{0mm}{0mm}Total stat. uncertainty   & 0.0027 	& $0.0091$ 	& $0.0049$ 	& $0.0034$ 	& $^{+0.10}_{-0.17}$
		& $^{+0.14}_{-0.15}$ 		& $0.049$ 	& 0.019          &         	$^{+0.055}_{-0.057}$\\[0.1ex]       \hline
      Mass factorisation                        & --            & 0.0007	& 0.0031 	& 0.0064	& 0.05 	& 0.05 	        & 0.002	         & 0.001   	& 0.004               \\
      Signal weights (stat.)                    & 0.0001        & 0.0001	& -- 	        & 0.0001	& --
      & --	& --	         & --   	& --               \\        
      $b$-hadron background      		& 0.0001	& 0.0004 	& 0.0004	& 0.0002 	& 0.02 	        & 0.02 		& 0.002 	& 0.003         & 0.001	\\ 
      $B_c^+$ feed-down			& 0.0005	& -- 		& -- 		& --		& --		&-- 			& --		& -- 		&	--		\\ 
      Angular resolution bias		& --	 	& -- 		& 0.0006 	& 0.0001 	& $^{+0.02}_{-0.03}$		 	& 0.01 		& -- 		& -- 	&	--		\\ 
      Ang. efficiency (reweighting)			& 0.0001 	& --            & 0.0011	& 0.0020	& 0.01 	        & --		& 0.001		& 0.005	        & 0.002            	\\ 
      Ang. efficiency (stat.) 			& 0.0001  	& 0.0002	& 0.0011	& 0.0004	& 0.02		& 0.01		& 0.004		& 0.002	        & 0.001	                \\ 
      Decay-time resolution       			& --     	& --    	& -- 		& -- 		& -- 	        & 0.01	& 0.002 	& 0.001 	& 0.005	\\ 
      Trigger efficiency (stat.)	& 0.0011 	& 0.0009 	& -- 		& -- 		& -- 		& -- 		& -- 		& --  		& --	\\ 
      Track reconstruction (simul.)			&  0.0007           & 0.0029	& 0.0005 	& 0.0006	 & $^{+0.01}_{-0.02}$ 	& 0.002 	        & 0.001	         & 0.001   	& $0.006$               \\
      Track reconstruction (stat.)			&  0.0005           & 0.0002	& -- 	& --	 & -- 	& -- 	        & --	         & --   	& 0.001               \\
      Length and momentum scales               	& 0.0002 	& -- 		& -- 		& -- 		& --	 	& -- 		& -- 		& -- 		& 0.005	\\ 
      S-P coupling factors	                & --            & --           & --            & --            & 0.01        & 0.01         & --             & 0.001       & 0.002            \\
      Fit bias 					& -- 		& -- 		& 0.0005 	& -- 		& -- 		&0.01 		& --		& 0.001 	&	--		\\ \hline
      \rule[3.5mm]{0mm}{0mm}Quadratic sum of syst.    		& 0.0015 	 & 	0.0032 	 & 	0.0036 	 & 	0.0067 	 & 	$^{+0.06}_{-0.07}$ 	 & 	0.06 	 & 	0.006 	 & 	0.007 	 & 	0.011	\\ 
      \hline
      \end{tabular}
  \end{center}
  \label{tab:finalsystematicssummary}
 	 
\end{table}

}

\begin{table}[tb]
  \caption{\small Statistical and systematic uncertainties for the polarisation-dependent result.}
\begin{center}\footnotesize
\begin{tabular} {l c c c c c c c c }
Source & $|\lambda^0|$ & $ |\lambda^{||}/\lambda^0|$ & $|\lambda^{\perp}/\lambda^0|$  &$|\lambda^{\rm S}/\lambda^0|$ & $\phi_s^{0}$& $\phi_s^{||}-\phi_s^{0}$& $\phi_s^{\perp}-\phi_s^{0}$ & $\phi_s^{\rm S}-\phi_s^{0}$\\
 &  &  &  &  &  [rad] &[rad]  & [rad] & [rad] \\
\hline
Total stat. uncertainty & 0.058 & 0.12 & 0.16 & 0.12 & 0.053 & 0.043 & 0.035 & 0.061 \\
\hline
Mass factorisation 		& 0.010 & 0.04 & 0.01 & 0.03 & 0.003 & 0.005 & 0.003 & 0.016 \\
$b$-hadron background 	&  0.002 & 0.01 & -- & 0.01 & 0.003 & 0.001 & 0.001 & 0.009 \\
Ang. efficiency (reweighting) & -- 	& -- 	& -- & 0.02 & 0.001 & 0.002 & 0.001 & 0.007 \\
Ang. efficiency (stat.) 		&  0.004 & 0.02 & 0.01 & 0.01 & 0.004 & 0.007 & 0.005 & 0.004 \\
Decay-time resolution 	&  0.006 & 0.01 & -- & 0.01 & 0.003 & 0.002 & 0.001 & 0.002 \\
S-P coupling factors 		& -- & -- & -- & -- & -- & -- & -- & 0.006 \\
\hline
      Quadratic sum of syst.    		&  0.013 & 0.05 & 0.01 & 0.04 & 0.007 & 0.009 & 0.006  & 	0.021	\\ 
\hline
\end{tabular}
\end{center}
\label{tab:finalsystematicssummary_pol}
\end{table}

A summary of systematic uncertainties is reported in Tables~\ref{tab:finalsystematicssummary}
and~\ref{tab:finalsystematicssummary_pol}.
The tagging parameters are constrained in the fit and therefore their associated systematic uncertainties
contribute to the statistical uncertainty of each parameter
in Table~\ref{tab:results}. This contribution is 0.004\rad to the statistical uncertainty on
$\phis$; 0.004\invps to that of $\dms$; 0.01\rad to that of $\delpar$ and is negligible for all other parameters.

The assumption that the $m(\jpsi K^{+}K^{-})$ distribution is independent from the decay time and angles is tested
by re-evaluating the signal weights in bins of the decay time and angles and repeating the fit. The
difference in fit results is assigned as a systematic uncertainty.
The systematic effect from the statistical uncertainty on the signal weights is determined by
re-computing them after varying the parameters of the $m(\jpsi K^{+}K^{-})$ fit model within
their statistical uncertainties, and assigning the difference in fit results as a systematic uncertainty. 

The effect due to the $b$-hadron background contributions is evaluated by varying
the proportion of simulated background events included in the fit by one standard deviation of their measured
fractions. In addition, a further systematic uncertainty is
assigned as the difference between the results of the fit to weighted or non-weighted data.

A small fraction of $\Bs\to\jpsi\Kp\Km$  decays come from the decays of $B_c^+$
mesons~\cite{LHCb-PAPER-2013-044}.
The effect of ignoring this component in the fit is evaluated using simulated pseudoexperiments where
a $0.8\%$ contribution~\cite{LHCb-PAPER-2013-044,Kiselev:2003mp}
of \Bs-from-$B_c^+$ decays is added
from a simulated sample of $B_c^+\to\Bs(\to \jpsi\phi)\pi^+$ decays.
Neglecting the $B_c^+$ component leads to a bias on
$\Gamma_s$ of $0.0005$\invps, which is added as a systematic uncertainty.
Other parameters are unaffected.

The decay angle resolution is found to be of the order of 20 mrad in simulated events.
The result of pseudoexperiments shows that ignoring this effect in the fit only leads to 
small biases in the polarisation amplitudes, which are assigned as systematic uncertainties.

The angular efficiency correction is determined from simulated
signal events weighted as in Ref.~\cite{LHCb-PAPER-2013-002} such that the
kinematic distributions of the final state particles match those in the data.
A systematic uncertainty is assigned as the difference between the fit results using angular corrections
from weighted or non-weighted simulated events. The limited size of the simulated sample leads to an additional
systematic uncertainty.

The systematic uncertainty from the 
decay time resolution parameters is not included in the statistical uncertainty of each parameter
and is now quoted explicitly. It is assigned as the difference of fit parameters obtained from the nominal
fit and a fit where the resolution model parameters are calibrated using
a sample of simulated prompt-$\jpsi$ events.

The trigger decay-time efficiency model, described in Ref.~\cite{LHCb-PAPER-2013-002}, 
introduces a systematic uncertainty that is determined by fixing
the value of each model parameter in the fit and subsequently repeating the fit with the parameter
values constrained within their statistical uncertainty. The quadratic differences
of the uncertainties returned by each fit are then assigned as systematic uncertainties.
The systematic effect of the track reconstruction efficiency is evaluated by applying the same
techniques on a large simulated sample of $\Bs\to\jpsi\phi$ decays. The differences between the generation and fitted values of each physics parameter in this sample is assigned as
the systematic uncertainty. The limited size of the control sample used to determine the track reconstruction
efficiency parameterisation leads to an additional systematic uncertainty.

The uncertainty on the longitudinal
coordinate of the LHCb vertex detector is found from survey data and
leads to an uncertainty on $\Gs$ and $\DGs$ of 0.020\%, with other parameters
being unaffected.
The momentum scale uncertainty is at most 0.022\%~\cite{LHCb-PAPER-2013-006}, which only affects \dms.

Different models of the S-wave line-shape and $m(\Kp\Km)$ resolution
are used to evaluate the
coupling factors in each of the six $m(\Kp\Km)$ bins and the resulting variation of the fit parameters are assigned as
systematic uncertainties. 
Possible biases of the fitting procedure are studied by generating and fitting many
simulated pseudoexperiments of equivalent size to the data. The resulting
biases are small, and those that are significantly different from zero
are assigned as systematic uncertainties.

The systematic correlations between parameters are evaluated by assuming that parameters
are fully (anti)correlated when the systematic uncertainty 
is determined by comparing results obtained from the nominal and a modified fit.
Other sources of systematic uncertainty
are assumed to have negligible parameter correlations. The combined
statistical and systematic correlation matrix is given in Ref.~\cite{supplementary}.

A measurement of $\phis$ and  $\maglambda$ by \lhcb using $\BtoJpsipipi$ decays 
of \mbox{$\phi_s^{\pi\pi} = 0.070 \pm 0.068 \pm 0.008 \rad$} and $|\lambda^{\pi\pi}| = 0.89 \pm 0.05 \pm 0.01$,
consistent with the measurement reported here, was reported in Ref.~\cite{LHCb-PAPER-2014-019}. 
The results from the two analyses are combined 
by incorporating the $\Bs\to \jpsi \Kp\Km$ result as a correlated Gaussian constraint
in the $\Bs\to \jpsi \pi^+\pi^-$ fit, 
under the assumption that $\Bs\to \jpsi \pi^+\pi^-$ and $\Bs\to \jpsi \Kp\Km$ decays
both proceed dominantly via $b\to c\overline{c}s$ transitions and the ratio between loop-induced processes and tree
diagrams are the same in each mode. 
The fit accounts for correlations between common parameters
and correlations between systematic uncertainties.
The combined result is $\phi_s = -0.010 \pm 0.039$~\rad and 
$|\lambda| = 0.957\pm 0.017$.
The correlation between the parameters is  about $-0.02$.

In conclusion, the \CP-violating phase $\phis$, and the $\Bs$ decay width parameters $\Gs$ and $\DGs$,
are measured using $\BtoJpsiKK$ decays selected from the full LHCb data set
from the first LHC operation period. The results are
$\phi_s = -0.058 \pm 0.049 \pm 0.006 \rad$,
$|\lambda| = 0.964\pm 0.019 \pm 0.007$,
$\Gamma_s = 0.6603  \pm 0.0027 \pm  0.0015 \invps$ and
$\DGs  =  0.0805    \pm  0.0091     \pm 0.0032  \invps$.
The parameter $\maglambda$ is consistent with unity, implying no evidence for \CP violation in $\Bs\to\jpsi\Kp\Km$ decays.
For the first time, the polarisation-dependent \CP-violating parameters are
measured and show no significant difference between the
four polarisation states.
The measurements of $\phi_s$ and $\maglambda$ in $\Bs\to \jpsi \Kp\Km$ decays
are consistent with those measured in $\BtoJpsipipi$ decays, 
and the combined results are  $\phi_s = -0.010\pm0.039$\rad and  $|\lambda| = 0.957\pm 0.017$.
The measurement of the \CP violating phase \phis is
the most precise to date and is in agreement with the SM prediction~\cite{CKMfitter},
in which it is assumed that sub-leading contributions to the decay amplitude are negligible.

We express our gratitude to our colleagues in the CERN
accelerator departments for the excellent performance of the LHC. We
thank the technical and administrative staff at the LHCb
institutes. We acknowledge support from CERN and from the national
agencies: CAPES, CNPq, FAPERJ and FINEP (Brazil); NSFC (China);
CNRS/IN2P3 (France); BMBF, DFG, HGF and MPG (Germany); SFI (Ireland); INFN (Italy); 
FOM and NWO (The Netherlands); MNiSW and NCN (Poland); MEN/IFA (Romania); 
MinES and FANO (Russia); MinECo (Spain); SNSF and SER (Switzerland); 
NASU (Ukraine); STFC (United Kingdom); NSF (USA).
The Tier1 computing centres are supported by IN2P3 (France), KIT and BMBF 
(Germany), INFN (Italy), NWO and SURF (The Netherlands), PIC (Spain), GridPP 
(United Kingdom).
We are indebted to the communities behind the multiple open 
source software packages on which we depend. We are also thankful for the 
computing resources and the access to software R\&D tools provided by Yandex LLC (Russia).
Individual groups or members have received support from 
EPLANET, Marie Sk\l{}odowska-Curie Actions and ERC (European Union), 
Conseil g\'{e}n\'{e}ral de Haute-Savoie, Labex ENIGMASS and OCEVU, 
R\'{e}gion Auvergne (France), RFBR (Russia), XuntaGal and GENCAT (Spain), Royal Society and Royal
Commission for the Exhibition of 1851 (United Kingdom).

\addcontentsline{toc}{section}{References}
\setboolean{inbibliography}{true}
\bibliographystyle{LHCb}
\bibliography{main,LHCb-PAPER,LHCb-CONF,LHCb-DP}

\ifx\mcitethebibliography\mciteundefinedmacro
\PackageError{LHCb.bst}{mciteplus.sty has not been loaded}
{This bibstyle requires the use of the mciteplus package.}\fi
\providecommand{\href}[2]{#2}
\begin{mcitethebibliography}{10}
\mciteSetBstSublistMode{n}
\mciteSetBstMaxWidthForm{subitem}{\alph{mcitesubitemcount})}
\mciteSetBstSublistLabelBeginEnd{\mcitemaxwidthsubitemform\space}
{\relax}{\relax}

\bibitem{Kobayashi:1973fv}
M.~Kobayashi and T.~Maskawa, \ifthenelse{\boolean{articletitles}}{{\it {$\CP$
  violation in the renormalizable theory of weak interaction}}, }{}Prog.\
  Theor.\ Phys.\  {\bf 49} (1973) 652\relax
\mciteBstWouldAddEndPuncttrue
\mciteSetBstMidEndSepPunct{\mcitedefaultmidpunct}
{\mcitedefaultendpunct}{\mcitedefaultseppunct}\relax
\EndOfBibitem
\bibitem{Cabibbo:1963yz}
N.~Cabibbo, \ifthenelse{\boolean{articletitles}}{{\it {Unitary symmetry and
  leptonic decays}},
  }{}\href{http://dx.doi.org/10.1103/PhysRevLett.10.531}{Phys.\ Rev.\ Lett.\
  {\bf 10} (1963) 531}\relax
\mciteBstWouldAddEndPuncttrue
\mciteSetBstMidEndSepPunct{\mcitedefaultmidpunct}
{\mcitedefaultendpunct}{\mcitedefaultseppunct}\relax
\EndOfBibitem
\bibitem{CKMfitter}
J.~Charles {\em et~al.}, \ifthenelse{\boolean{articletitles}}{{\it {Predictions
  of selected flavour observables within the Standard Model}},
  }{}\href{http://dx.doi.org/10.1103/PhysRevD.84.033005}{Phys.\ Rev.\  {\bf
  D84} (2011) 033005}, \href{http://arxiv.org/abs/1106.4041}{{\tt
  arXiv:1106.4041}}, with updated results and plots available at
  \url{http://ckmfitter.in2p3.fr}\relax
\mciteBstWouldAddEndPuncttrue
\mciteSetBstMidEndSepPunct{\mcitedefaultmidpunct}
{\mcitedefaultendpunct}{\mcitedefaultseppunct}\relax
\EndOfBibitem
\bibitem{Buras:2009if}
A.~J. Buras, \ifthenelse{\boolean{articletitles}}{{\it {Flavour theory: 2009}},
  }{}PoS {\bf EPS-HEP2009} (2009) 024,
  \href{http://arxiv.org/abs/0910.1032}{{\tt arXiv:0910.1032}}\relax
\mciteBstWouldAddEndPuncttrue
\mciteSetBstMidEndSepPunct{\mcitedefaultmidpunct}
{\mcitedefaultendpunct}{\mcitedefaultseppunct}\relax
\EndOfBibitem
\bibitem{Chiang:2009ev}
C.-W. Chiang {\em et~al.}, \ifthenelse{\boolean{articletitles}}{{\it {New
  physics in $\Bs \to \jpsi\phi$: A general analysis}},
  }{}\href{http://dx.doi.org/10.1007/JHEP04(2010)031}{JHEP {\bf 04} (2010)
  031}, \href{http://arxiv.org/abs/0910.2929}{{\tt arXiv:0910.2929}}\relax
\mciteBstWouldAddEndPuncttrue
\mciteSetBstMidEndSepPunct{\mcitedefaultmidpunct}
{\mcitedefaultendpunct}{\mcitedefaultseppunct}\relax
\EndOfBibitem
\bibitem{cLHCb-PAPER-2011-021}
LHCb collaboration, R.~Aaij {\em et~al.},
  \ifthenelse{\boolean{articletitles}}{{\it {Measurement of the \CP-violating
  phase $\phi_s$ in the decay $B^0_s \to J/\psi \phi$}},
  }{}\href{http://dx.doi.org/10.1103/PhysRevLett.108.101803}{Phys.\ Rev.\
  Lett.\  {\bf 108} (2012) 101803}, \href{http://arxiv.org/abs/1112.3183}{{\tt
  arXiv:1112.3183}}\relax
\mciteBstWouldAddEndPuncttrue
\mciteSetBstMidEndSepPunct{\mcitedefaultmidpunct}
{\mcitedefaultendpunct}{\mcitedefaultseppunct}\relax
\EndOfBibitem
\bibitem{LHCb-PAPER-2013-002}
LHCb collaboration, R.~Aaij {\em et~al.},
  \ifthenelse{\boolean{articletitles}}{{\it {Measurement of $\CP$ violation and
  the $B^0_s$ meson decay width difference with $B_s^0\to J/\psi K^+K^-$ and
  $B_s^0 \to J/\psi\pi^+\pi^-$ decays}},
  }{}\href{http://dx.doi.org/10.1103/PhysRevD.87.112010}{Phys.\ Rev.\  {\bf
  D87} (2013) 112010}, \href{http://arxiv.org/abs/1304.2600}{{\tt
  arXiv:1304.2600}}\relax
\mciteBstWouldAddEndPuncttrue
\mciteSetBstMidEndSepPunct{\mcitedefaultmidpunct}
{\mcitedefaultendpunct}{\mcitedefaultseppunct}\relax
\EndOfBibitem
\bibitem{Aad:2014cqa}
ATLAS Collaboration, G.~Aad {\em et~al.},
  \ifthenelse{\boolean{articletitles}}{{\it {Flavour tagged time dependent
  angular analysis of the $B_s \rightarrow J/\psi \phi$ decay and extraction of
  $\Delta\Gamma_s$ and the weak phase $\phi_s$ in ATLAS}},
  }{}\href{http://dx.doi.org/10.1103/PhysRevD.90.052007}{Phys.\ Rev.\  {\bf
  D90} (2014) 052007}, \href{http://arxiv.org/abs/1407.1796}{{\tt
  arXiv:1407.1796}}\relax
\mciteBstWouldAddEndPuncttrue
\mciteSetBstMidEndSepPunct{\mcitedefaultmidpunct}
{\mcitedefaultendpunct}{\mcitedefaultseppunct}\relax
\EndOfBibitem
\bibitem{Aaltonen:2012ie}
CDF collaboration, T.~Aaltonen {\em et~al.},
  \ifthenelse{\boolean{articletitles}}{{\it {Measurement of the bottom-strange
  meson mixing phase in the full CDF data set}},
  }{}\href{http://dx.doi.org/10.1103/PhysRevLett.109.171802}{Phys.\ Rev.\
  Lett.\  {\bf 109} (2012) 171802}, \href{http://arxiv.org/abs/1208.2967}{{\tt
  arXiv:1208.2967}}\relax
\mciteBstWouldAddEndPuncttrue
\mciteSetBstMidEndSepPunct{\mcitedefaultmidpunct}
{\mcitedefaultendpunct}{\mcitedefaultseppunct}\relax
\EndOfBibitem
\bibitem{Abazov:2011ry}
D0 collaboration, V.~M. Abazov {\em et~al.},
  \ifthenelse{\boolean{articletitles}}{{\it {Measurement of the \CP-violating
  phase $\phi_s^{J/\psi \phi}$ using the flavor-tagged decay $B_s^0 \to J/\psi
  \phi$ in 8 fb$^{-1}$ of $p\bar{p}$ collisions}},
  }{}\href{http://dx.doi.org/10.1103/PhysRevD.85.032006}{Phys.\ Rev.\  {\bf
  D85} (2012) 032006}, \href{http://arxiv.org/abs/1109.3166}{{\tt
  arXiv:1109.3166}}\relax
\mciteBstWouldAddEndPuncttrue
\mciteSetBstMidEndSepPunct{\mcitedefaultmidpunct}
{\mcitedefaultendpunct}{\mcitedefaultseppunct}\relax
\EndOfBibitem
\bibitem{LHCb-PAPER-2014-019}
LHCb collaboration, R.~Aaij {\em et~al.},
  \ifthenelse{\boolean{articletitles}}{{\it {Measurement of the $\CP$-violating
  phase $\phi_s$ in $\overline{B}^0_s\to J/\psi\pi^+\pi^-$ decays}},
  }{}\href{http://dx.doi.org/10.1016/j.nuclphysb.2014.06.011}{Phys.\ Lett.\
  {\bf B736} (2014) 186}, \href{http://arxiv.org/abs/1405.4140}{{\tt
  arXiv:1405.4140}}\relax
\mciteBstWouldAddEndPuncttrue
\mciteSetBstMidEndSepPunct{\mcitedefaultmidpunct}
{\mcitedefaultendpunct}{\mcitedefaultseppunct}\relax
\EndOfBibitem
\bibitem{Faller:2008gt}
S.~Faller, R.~Fleischer, and T.~Mannel,
  \ifthenelse{\boolean{articletitles}}{{\it {Precision physics with $B^0_s \to
  J/\psi \phi$ at the LHC: the quest for new physics}},
  }{}\href{http://dx.doi.org/10.1103/PhysRevD.79.014005}{Phys.\ Rev.\  {\bf
  D79} (2009) 014005}, \href{http://arxiv.org/abs/0810.4248}{{\tt
  arXiv:0810.4248}}\relax
\mciteBstWouldAddEndPuncttrue
\mciteSetBstMidEndSepPunct{\mcitedefaultmidpunct}
{\mcitedefaultendpunct}{\mcitedefaultseppunct}\relax
\EndOfBibitem
\bibitem{Alves:2008zz}
LHCb collaboration, A.~A. Alves~Jr. {\em et~al.},
  \ifthenelse{\boolean{articletitles}}{{\it {The \lhcb detector at the LHC}},
  }{}\href{http://dx.doi.org/10.1088/1748-0221/3/08/S08005}{JINST {\bf 3}
  (2008) S08005}\relax
\mciteBstWouldAddEndPuncttrue
\mciteSetBstMidEndSepPunct{\mcitedefaultmidpunct}
{\mcitedefaultendpunct}{\mcitedefaultseppunct}\relax
\EndOfBibitem
\bibitem{LHCb-DP-2012-004}
R.~Aaij {\em et~al.}, \ifthenelse{\boolean{articletitles}}{{\it {The \lhcb
  trigger and its performance in 2011}},
  }{}\href{http://dx.doi.org/10.1088/1748-0221/8/04/P04022}{JINST {\bf 8}
  (2013) P04022}, \href{http://arxiv.org/abs/1211.3055}{{\tt
  arXiv:1211.3055}}\relax
\mciteBstWouldAddEndPuncttrue
\mciteSetBstMidEndSepPunct{\mcitedefaultmidpunct}
{\mcitedefaultendpunct}{\mcitedefaultseppunct}\relax
\EndOfBibitem
\bibitem{Stone:2008ak}
S.~Stone and L.~Zhang, \ifthenelse{\boolean{articletitles}}{{\it {S-waves and
  the measurement of \CP violating phases in $B_s^0$ decays}},
  }{}\href{http://dx.doi.org/10.1103/PhysRevD.79.074024}{Phys.\ Rev.\  {\bf
  D79} (2009) 074024}, \href{http://arxiv.org/abs/0812.2832}{{\tt
  arXiv:0812.2832}}\relax
\mciteBstWouldAddEndPuncttrue
\mciteSetBstMidEndSepPunct{\mcitedefaultmidpunct}
{\mcitedefaultendpunct}{\mcitedefaultseppunct}\relax
\EndOfBibitem
\bibitem{Santos:2013gra}
D.~Martinez~Santos and F.~Dupertuis, \ifthenelse{\boolean{articletitles}}{{\it
  {Mass distributions marginalized over per-event errors}},
  }{}\href{http://dx.doi.org/10.1016/j.nima.2014.06.081}{Nucl.\ Instrum.\
  Meth.\  {\bf A764} (2014) 150}, \href{http://arxiv.org/abs/1312.5000}{{\tt
  arXiv:1312.5000}}\relax
\mciteBstWouldAddEndPuncttrue
\mciteSetBstMidEndSepPunct{\mcitedefaultmidpunct}
{\mcitedefaultendpunct}{\mcitedefaultseppunct}\relax
\EndOfBibitem
\bibitem{LHCb-PAPER-2013-033}
LHCb collaboration, R.~Aaij {\em et~al.},
  \ifthenelse{\boolean{articletitles}}{{\it {Measurement of the
  flavour-specific $\CP$-violating asymmetry $a_{\rm sl}^s$ in $B_s^0$
  decays}}, }{}\href{http://dx.doi.org/10.1016/j.physletb.2013.12.030}{Phys.\
  Lett.\  {\bf B728} (2014) 607}, \href{http://arxiv.org/abs/1308.1048}{{\tt
  arXiv:1308.1048}}\relax
\mciteBstWouldAddEndPuncttrue
\mciteSetBstMidEndSepPunct{\mcitedefaultmidpunct}
{\mcitedefaultendpunct}{\mcitedefaultseppunct}\relax
\EndOfBibitem
\bibitem{sFit}
Y.~{Xie}, \ifthenelse{\boolean{articletitles}}{{\it {sFit: A method for
  background subtraction in maximum likelihood fit}},
  }{}\href{http://arxiv.org/abs/0905.0724}{{\tt arXiv:0905.0724}}\relax
\mciteBstWouldAddEndPuncttrue
\mciteSetBstMidEndSepPunct{\mcitedefaultmidpunct}
{\mcitedefaultendpunct}{\mcitedefaultseppunct}\relax
\EndOfBibitem
\bibitem{LHCb-PAPER-2013-065}
LHCb collaboration, R.~Aaij {\em et~al.},
  \ifthenelse{\boolean{articletitles}}{{\it {Measurements of the $B^+$, $B^0$,
  $B_s^0$ meson and $\Lambda_b^0$ baryon lifetimes}},
  }{}\href{http://dx.doi.org/10.1007/JHEP04(2014)114}{JHEP {\bf 04} (2014)
  114}, \href{http://arxiv.org/abs/1402.2554}{{\tt arXiv:1402.2554}}\relax
\mciteBstWouldAddEndPuncttrue
\mciteSetBstMidEndSepPunct{\mcitedefaultmidpunct}
{\mcitedefaultendpunct}{\mcitedefaultseppunct}\relax
\EndOfBibitem
\bibitem{LHCb-CONF-2012-033}
{LHCb collaboration}, \ifthenelse{\boolean{articletitles}}{{\it {Optimization
  and calibration of the same-side kaon tagging algorithm using hadronic $\Bs$
  decays in 2011 data}}, }{}
  \href{http://cdsweb.cern.ch/search?p={LHCb-CONF-2012-033}&f=reportnumber&action_search=Search&c=LHCb+Reports&c=LHCb+Conference+Proceedings&c=LHCb+Conference+Contributions&c=LHCb+Notes&c=LHCb+Theses&c=LHCb+Papers}
  {{LHCb-CONF-2012-033}}\relax
\mciteBstWouldAddEndPuncttrue
\mciteSetBstMidEndSepPunct{\mcitedefaultmidpunct}
{\mcitedefaultendpunct}{\mcitedefaultseppunct}\relax
\EndOfBibitem
\bibitem{supplementary}
See supplementary material for details.\relax
\mciteBstWouldAddEndPunctfalse
\mciteSetBstMidEndSepPunct{\mcitedefaultmidpunct}
{}{\mcitedefaultseppunct}\relax
\EndOfBibitem
\bibitem{LHCb-PAPER-2013-006}
LHCb collaboration, R.~Aaij {\em et~al.},
  \ifthenelse{\boolean{articletitles}}{{\it {Precision measurement of the
  $B^0_s-\overline{B}^0_s$ oscillation frequency in the decay $B^0_s \to D^+_s
  \pi^-$}}, }{}\href{http://dx.doi.org/10.1088/1367-2630/15/5/053021}{New J.\
  Phys.\  {\bf 15} (2013) 053021}, \href{http://arxiv.org/abs/1304.4741}{{\tt
  arXiv:1304.4741}}\relax
\mciteBstWouldAddEndPuncttrue
\mciteSetBstMidEndSepPunct{\mcitedefaultmidpunct}
{\mcitedefaultendpunct}{\mcitedefaultseppunct}\relax
\EndOfBibitem
\bibitem{Liu:2013nea}
X.~Liu, W.~Wang, and Y.~Xie, \ifthenelse{\boolean{articletitles}}{{\it {Penguin
  pollution in $B\to J/\psi V$ decays and impact on the extraction of the
  $B_s-\bar B_s$ mixing phase}},
  }{}\href{http://dx.doi.org/10.1103/PhysRevD.89.094010}{Phys.\ Rev.\  {\bf
  D89} (2014) 094010}, \href{http://arxiv.org/abs/1309.0313}{{\tt
  arXiv:1309.0313}}\relax
\mciteBstWouldAddEndPuncttrue
\mciteSetBstMidEndSepPunct{\mcitedefaultmidpunct}
{\mcitedefaultendpunct}{\mcitedefaultseppunct}\relax
\EndOfBibitem
\bibitem{LHCb-PAPER-2013-044}
LHCb collaboration, R.~Aaij {\em et~al.},
  \ifthenelse{\boolean{articletitles}}{{\it {Observation of the decay $B_c^+\to
  B_s^0\pi^+$}},
  }{}\href{http://dx.doi.org/10.1103/PhysRevLett.111.181801}{Phys.\ Rev.\
  Lett.\  {\bf 111} (2013) 181801}, \href{http://arxiv.org/abs/1308.4544}{{\tt
  arXiv:1308.4544}}\relax
\mciteBstWouldAddEndPuncttrue
\mciteSetBstMidEndSepPunct{\mcitedefaultmidpunct}
{\mcitedefaultendpunct}{\mcitedefaultseppunct}\relax
\EndOfBibitem
\bibitem{Kiselev:2003mp}
V.~Kiselev, \ifthenelse{\boolean{articletitles}}{{\it {Decays of the $B_c$
  meson}}, }{}\href{http://arxiv.org/abs/hep-ph/0308214}{{\tt
  arXiv:hep-ph/0308214}}\relax
\mciteBstWouldAddEndPuncttrue
\mciteSetBstMidEndSepPunct{\mcitedefaultmidpunct}
{\mcitedefaultendpunct}{\mcitedefaultseppunct}\relax
\EndOfBibitem
\end{mcitethebibliography}

\newpage
\section{Supplementary material for PRL}
\label{sec:supplementary}

\begin{table}[h!]
\caption{\small Statistical correlation matrix from the polarisation-independent fit.}
\begin{center}\small
\begin{tabular}{l c c c c c c c c c} 
 & $\Gamma_s$ & $\DGs$ & $|A_\perp|^2$ & $|A_0|^2$ & $\delta_\parallel$ & $\delta_\perp$ & $\phi_s$ & $|\lambda|$ & $\Delta m_s$ \\\hline
$\Gamma_s$ & ${+1.00}$ & ${-0.45}$ & ${+0.39}$ & ${-0.31}$ & $-0.07$ & $-0.02$ & $+0.01$ & $-0.01$ & $+0.01$\\
$\DGs$ &  & ${+1.00}$ & ${-0.69}$ & ${+0.65}$ & $+0.02$ & $-0.03$ & $-0.08$ & $+0.02$ & $-0.03$\\
$|A_\perp|^2$ &  &  & ${+1.00}$ & ${-0.59}$ & $-0.29$ & $-0.10$ & $+0.04$ & $-0.03$ & $+0.00$\\
$|A_0|^2$ &  &  &  & ${+1.00}$ & $-0.02$ & $-0.04$ & $-0.03$ & $+0.02$ & $-0.03$\\
$\delta_\parallel$ &  &  &  &  & ${+1.00}$ & ${+0.42}$ & $+0.01$ & $+0.05$ & $+0.05$\\
$\delta_\perp$ &  &  &  &  &  & ${+1.00}$ & $+0.14$ & $-0.17$ & ${+0.67}$\\
$\phi_s$ &  &  &  &  &  &  & ${+1.00}$ & $-0.02$ & $+0.09$\\
$|\lambda|$ &  &  &  &  &  &  &  & ${+1.00}$ & $-0.21$\\
$\Delta m_s$ &  &  &  &  &  &  &  &  & ${+1.00}$\\

 \end{tabular}
\end{center}
\label{tab:results-corr}
\end{table}

\begin{table}[h!]
\caption{\small Total statistical and systematic correlation matrix from the polarisation-independent fit.}
\begin{center}\small
\begin{tabular}{l c c c c c c c c c} 
	&	$\Gamma_s$ 	 & $\Delta\Gamma_s$ 	 & $|A_\perp|^2$ 	 & $|A_0|^2$ 	 & $\delta_\parallel$ 	 & $\delta_\perp$ 	 & $\phi_s$ 	 & $|\lambda|$ 	 & $\Delta m_s $ 	 \\
\hline
$\Gamma_s$ &  ${+1.00}$ & ${-0.30}$ & $+0.25$ & $-0.09$ & $-0.08$ & $-0.01$ & $+0.01$ & $-0.01$ & $+0.03$  \\
$\Delta\Gamma_s$ &   &  ${+1.00}$ & ${-0.59}$ & ${+0.36}$ & $-0.06$ & $-0.05$ & $-0.08$ & $+0.03$ & $-0.00$  \\
$|A_\perp|^2$ &   &   &  ${+1.00}$ & ${-0.70}$ & $+0.03$ & $+0.08$ & $+0.05$ & $+0.00$ & $+0.02$  \\
$|A_0|^2$ &   &   &   &  ${+1.00}$ & ${-0.38}$ & $-0.28$ & $-0.04$ & $-0.01$ & $-0.05$  \\
$\delta_\parallel$ &   &   &   &   &  ${+1.00}$ & ${+0.47}$ & $+0.03$ & $+0.00$ & $+0.05$  \\
$\delta_\perp$ &   &   &   &   &   &  ${+1.00}$ & $+0.15$ & $-0.18$ & ${+0.64}$  \\
$\phi_s$ &   &   &   &   &   &   &  ${+1.00}$ & $-0.03$ & $+0.09$  \\
$|\lambda|$ &   &   &   &   &   &   &   &  ${+1.00}$ & $-0.21$  \\
$\Delta m_s$ &   &   &   &   &   &   &   &   &  ${+1.00}$  \\
 \end{tabular}
\end{center}
\label{tab:results-corr-tot}
\end{table}

 \begin{table}[h!]
\caption{\small Statistical correlation matrix for the $\phi_s^{k}$ and $|\lambda^k|$  parameters from the polarisation-dependent fit.}
\centerline{\small
\begin{tabular}{ccccccccc}
& $|\lambda^0|$ & $|\lambda^{\parallel}/\lambda^0|$ & $|\lambda^{\perp}/\lambda^0|$ & $|\lambda^{\rm S}/\lambda^0|$ &  $\phi^{0}_s$ &  $\phi_s^{\parallel}-\phi_s^{0}$ &  $\phi_s^{\perp}-\phi_s^0$ &  $\phi_s^{\rm S}-\phi_s^0$ \\
\hline
$|\lambda^{0}|$ &  ${+1.00}$ & ${-0.32}$ & ${-0.59}$ & ${-0.89}$ & $+0.01$ & $-0.08$ & $-0.06$ & $+0.00$  \\
$|\lambda^{\parallel}/\lambda^0|$ &   &  ${+1.00}$ & $-0.23$ & $+0.27$ & $+0.00$ & ${+0.31}$ & $+0.16$ & $+0.10$  \\
$|\lambda^{\perp}/\lambda^0|$ &   &   &  ${+1.00}$ & ${+0.53}$ & $-0.02$ & $+0.06$ & $-0.29$ & $-0.02$  \\
$|\lambda^{\rm S}/\lambda^0|$ &   &   &   &  ${+1.00}$ & $-0.01$ & $+0.07$ & $+0.06$ & $+0.22$  \\
$\phi^{0}_s$ &   &   &   &   &  ${+1.00}$ & $-0.14$ & $+0.13$ & $+0.14$  \\
$\phi_s^{\parallel}-\phi_s^{0}$ &   &   &   &   &   &  ${+1.00}$ & ${+0.52}$ & $+0.11$  \\
$\phi_s^{\perp}-\phi_s^0$ &   &   &   &   &   &   &  ${+1.00}$ & $+0.08$  \\
$\phi_s^{\rm S}-\phi_s^0$ &   &   &   &   &   &   &   &  ${+1.00}$  \\
\end{tabular}
}
\label{tab:phipol-correlation}
\end{table}

\begin{table}[h!]
\caption{\small Total statistical and systematic correlation matrix for the $\phi_s^{k}$ and $|\lambda^k|$ parameters from the polarisation-dependent fit.}
\centerline{\small
\begin{tabular}{ccccccccc}
& $|\lambda^0|$ & $|\lambda^{\parallel}/\lambda^0|$ & $|\lambda^{\perp}/\lambda^0|$ & $|\lambda^{\rm S}/\lambda^0|$ &  $\phi^{0}_s$ &  $\phi_s^{\parallel}-\phi_s^{0}$ &  $\phi_s^{\perp}-\phi_s^0$ &  $\phi_s^{\rm S}-\phi_s^0$ \\
\hline
$|\lambda^{0}|$ &  ${+1.00}$ & ${-0.35}$ & ${-0.56}$ & ${-0.77}$ & $-0.00$ & $-0.09$ & $-0.07$ & $-0.02$  \\
$|\lambda^{\parallel}/\lambda^0|$ &   &  ${+1.00}$ & $-0.23$ & $+0.15$ & $+0.02$ & ${+0.32}$ & $+0.17$ & $+0.14$  \\
$|\lambda^{\perp}/\lambda^0|$ &   &   &  ${+1.00}$ & ${+0.51}$ & $-0.02$ & $+0.05$ & $-0.29$ & $-0.03$  \\
$|\lambda^{\rm S}/\lambda^0|$ &   &   &   &  ${+1.00}$ & $-0.03$ & $+0.03$ & $+0.04$ & $+0.18$  \\
$\phi^{0}_s$ &   &   &   &   &  ${+1.00}$ & $-0.13$ & $+0.13$ & $+0.14$  \\
$\phi_s^{\parallel}-\phi_s^{0}$ &   &   &   &   &   &  ${+1.00}$ & ${+0.51}$ & $+0.12$  \\
$\phi_s^{\perp}-\phi_s^0$ &   &   &   &   &   &   &  ${+1.00}$ & $+0.09$  \\
$\phi_s^{\rm S}-\phi_s^0$ &   &   &   &   &   &   &   &  ${+1.00}$  \\
\end{tabular}
}
\label{tab:results-corr-tot-pol-dep}
\end{table}

\begin{table}[htbp]     
  \centering
  \caption{\small S-wave parameter estimates in each $m(\Kp\Km)$ bin from the polarisation-independent fit. 
  Only statistical uncertainties are shown.}
	\label{tab:Swave_parameters}
  \begin{tabular}{lc}
    Parameter                               &  Value       \\
    \hline

    $F_\text{S1}$                                   &  $0.426\phantom{0}  \pm 0.054\phantom{0} $   \\
    $F_\text{S2}$                                   &  $0.059\phantom{0}  \pm 0.017\phantom{0} $   \\
    $F_\text{S3}$                                   &  $0.0101 \pm 0.0067$   \\
    $F_\text{S4}$                                   &  $0.0103 \pm 0.0061$   \\
    $F_\text{S5}$                                 &  $0.049\phantom{0}  \pm 0.015\phantom{0} $   \\
    $F_\text{S6}$                                   &  $0.193\phantom{0}  \pm 0.025\phantom{0} $   \\
    \hline \\[-2.5ex]

    $\delta_\text{S1}-\delta_\perp$                &  $\phantom{+}0.84   \pm 0.20\phantom{0}\,  $  \\
    $\delta_\text{S2}-\delta_\perp$                &  $\phantom{+}2.15   \pm 0.28\phantom{0}\,  $  \\
    $\delta_\text{S3}-\delta_\perp$                 &  $\phantom{+}0.47   \pm 0.21\phantom{0}\,  $  \\
    $\delta_\text{S4}-\delta_\perp$                &  $-0.34   \pm 0.17\phantom{0}\,  $  \\
    $\delta_\text{S5}-\delta_\perp$                 &  $-0.59   \pm 0.15\phantom{0}\,  $  \\
    $\delta_\text{S6}-\delta_\perp$                 &  $-0.90   \pm 0.14\phantom{0}\,  $  \\
    \hline
  \end{tabular}
 
\end{table}

\clearpage

\newpage
\centerline{\large\bf LHCb collaboration}
\begin{flushleft}
\small
R.~Aaij$^{41}$, 
B.~Adeva$^{37}$, 
M.~Adinolfi$^{46}$, 
A.~Affolder$^{52}$, 
Z.~Ajaltouni$^{5}$, 
S.~Akar$^{6}$, 
J.~Albrecht$^{9}$, 
F.~Alessio$^{38}$, 
M.~Alexander$^{51}$, 
S.~Ali$^{41}$
% ...

{\footnotesize \it
% ...
$ ^{5}$Clermont Universit\'{e}, Universit\'{e} Blaise Pascal, CNRS/IN2P3, LPC, Clermont-Ferrand, France\\
$ ^{6}$CPPM, Aix-Marseille Universit\'{e}, CNRS/IN2P3, Marseille, France\\
% ...
$ ^{9}$Fakult\"{a}t Physik, Technische Universit\"{a}t Dortmund, Dortmund, Germany\\
% ...
$ ^{37}$Universidad de Santiago de Compostela, Santiago de Compostela, Spain\\
$ ^{38}$European Organization for Nuclear Research (CERN), Geneva, Switzerland\\
% ...
$ ^{41}$Nikhef National Institute for Subatomic Physics, Amsterdam, The Netherlands\\
% ...
$ ^{46}$H.H. Wills Physics Laboratory, University of Bristol, Bristol, United Kingdom\\
% ...
$ ^{51}$School of Physics and Astronomy, University of Glasgow, Glasgow, United Kingdom\\
$ ^{52}$Oliver Lodge Laboratory, University of Liverpool, Liverpool, United Kingdom\\
% ...
}
\end{flushleft}

\end{document}